\documentclass[12pt,journal,draftclsnofoot,onecolumn]{IEEEtran}
\usepackage[latin9]{inputenc}
\usepackage{amsmath}
\usepackage{amssymb}
\usepackage{graphicx}
\usepackage{esint}

\makeatletter

\providecommand{\tabularnewline}{\\}

\@ifundefined{date}{}{\date{}}

\usepackage{amsthm}\usepackage{amsthm}

\makeatother

\begin{document}

\title{A Subset Selection Algorithm for Wireless Sensor Networks}

\author{\IEEEauthorblockN{Seyed~Hamed~Mousavi\IEEEauthorrefmark{1},~
Javad~Haghighat\IEEEauthorrefmark{1},~Walaa~Hamouda\IEEEauthorrefmark{2},
~\emph{Senior Member,~IEEE}, and Reza~Dastbasteh\IEEEauthorrefmark{3}}
\thanks{\IEEEauthorrefmark{1}Department of Electrical Engineering, Shiraz
University of Technology, Iran (Email: {s.moosavi,haghighat}@sutech.ac.ir)%
}
\thanks{\IEEEauthorrefmark{2}Electrical and Computer Engineering, Concordia
University, Montreal, QC, Canada, H3G 1M8 (Email: hamouda@ece.concordia.ca)%
}
\thanks{\IEEEauthorrefmark{3}Department of Mathematics, Shiraz University,
Iran (Email: r.dastbasteh@shirazu.ac.ir)%
}}
\maketitle
\begin{abstract}
One of the main challenges facing wireless sensor networks (WSNs)
is the limited power resources available at small sensor nodes. It
is therefore desired to reduce the power consumption of sensors while
keeping the distortion between the source information and its estimate
at the fusion centre (FC) below a specific threshold. In this paper,
given the channel state information at the FC, we propose a subset
selection algorithm of sensor nodes to reduce the average transmission
power of the WSN. We assume the channels between the source and the
sensors to be correlated fading channels, modeled by the Gilbert-Elliott
model. We show that when these channels are known at the FC, a subset
of sensors can be selected by the FC such that the received observations
from this subset is sufficient to estimate the source information
at the FC while maintaining the distortion between source information
and its estimate below a specific threshold. Through analyses, we
find the probability distribution of the size of this subset and provide
results to evaluate the power efficiency of our proposed algorithm. \end{abstract}

\section{\label{sec:Introduction}Introduction}

Wireless sensor networks (WSNs) \cite{Survey_WSN}, \cite{singlenode1-1,singlenode2-1,Jointpoweralloc}
are receiving increasing attention due their numerous current and
foreseen applications in several fields including field trials and
performance monitoring of solar panels \cite{sola_ WSN}, target detection
through digital cameras \cite{target_WSN}, and petrochemical industry
fields \cite{petro_WSN}. One of the main challenges of WSNs is to
overcome their energy constraint problem. That is, sensors are powered
by batteries with limited energy budgets. Due to deployment of sensor
nodes in inaccessible or hostile environments, recharging these batteries
is often not an option. Also, the network is expected to have a lifetime
in the order of several months, or even years \cite{Energy_Survey_WSN}.
Therefore, design of power-efficient WSNs is of extreme importance.
Numerous works are dedicated to the topic of energy conservation in
WSNs. A comprehensive review of these works is given in \cite{Energy_Survey_WSN}.

A typical sensor node in a WSN consists of three main subsystems namely;
sensing, processing, and wireless communication subsystems. These
subsystems are responsible for data acquisition, local data processing,
and data transmission, respectively \cite{Energy_Survey_WSN}. In
addition, a power source is included in the sensor node with a limited
power budget. In \cite{Energy_aware_microsensor}, it is shown using
experimental measurements that in most cases data transmission is
the most energy-consuming unit of the sensor node. In a similar conclusion,
it is estimated in \cite{WSN_integrated} that transmitting one bit
by the data communication unit requires an energy equivalent to performing
about a thousand operations in the data processing unit. It is worth
mentioning that in some applications, the sensing subsystem might
consume more power than the data communication subsystem (see \cite{Energy_Survey_WSN}
for details) but in typical applications of WSNs, the highest portion
of power is consumed by the data communication subsystem. Therefore,
it is highly desired to develop protocols to reduce the transmission
power of sensor nodes and hence extend their lifetime.

In this paper, we consider a WSN where the source-sensor channels
are correlated fading channels modeled as Gilbert-Elliott channels
\cite{Gilbert}, \cite{Elliott} . The Gilbert-Elliott model is a
first-order Markov model for a correlated fading channel quantized
to binary levels of Good and Bad states by setting a proper Signal-to-Noise
Ratio (SNR) threshold. The channel in its Good and Bad states is modeled
by binary-symmetric channels (BSCs) with crossover probabilities of
$p_{G}$ and $p_{B}$, respectively. The Gilbert-Elliott model is
the simplest possible finite-state Markov (FSM) model for correlated
fading channels. The problem of modeling a correlated fading channel
by a FSM process is considered in numerous works. An excellent review
of works on FSM modeling of fading channels is provided in \cite{Sadeghi_Fading_Review}
where the relations between real-valued fading channel parameters
and the FSM channel parameters are also considered.

Let a binary source block consisting of $M$ bits be transmitted to
$N$ sensors via independent Gilbert-Eliott channels. For each source-sensor
channel, the channel states during this transmission can be expressed
as an $M$ bit binary sequence where we let a bit $1$ represent a
Good state and a bit $0$ represent a Bad state. We call this $M$-bit
sequence as the channel-state information Sequence (CSI sequence).
For slowly varying fading channels the CSI sequence consists of a
few runs and is efficiently compressed using a run-length code (See
Fig. \ref{fig:Realization-of-Gilbert-Elliott} and Table \ref{tab:fdTs-epsilon-mu-R}).

Our main contribution in this paper is to propose and analyze a two-phase
transmission scheme as follows. At the first phase, each sensor compresses
its respective source-sensor CSI using a run-length code and transmits
it to the FC. Based on the received CSI from all nodes, the FC will
know the location of Good bits, i.e. the bits that are received in
a Good channel state. The FC then finds the smallest subset of sensors
such that for each source bit, at least one of the sensors in the
subset has a Good observation of that source bit. In other words,
this is the subset with minimum number of sensors, such that for each
source bit at least one of the sensors received this bit in Good channel
state. Then, the FC sends a feedback signal to request transmission
from this subset. Therefore, at the second phase, only a subset of
sensors transmit to the FC, resulting in reduction in the average
transmission power.

The motivation behind our proposed algorithm is as follows. According
to the Gilbert-Elliott model \cite{Gilbert,Elliott}, we have $p_{G}<p_{B}$,
i.e. Good bits are more reliable than Bad bits. Therefore, we are
in fact attempting to find the minimum number of sensors such that
if these sensors transmit to the FC and the rest of sensors remain
silent, the FC still receives one (or more than one) reliable copy
of each source bit and consequently is able to reliably reconstruct
the source information. To examine this idea more precisely, let the
WSN have an end-to-end distortion requirement of $D\leq\hat{D}$,
where $D$ is the expected value of the normalized Hamming distortion
(the Bit Error Rate) between the source and its estimate at the FC;
and $\hat{D}$ is a fixed distortion threshold. If a (minimum-sized)
subset of sensors exists such that each source bit is received through
a Good channel by at least one of the sensors in the subset, then
the FC will be able to reconstruct the source with a distortion less
than or equal $p_{G}$. Let $\nu$ be the probability of existence
of such subset. Then we could bound the end-to-end distortion of the
WSN as $D\leq D_{u}$ where $D_{u}=\nu\times p_{G}+\left(1-\nu\right)\times\frac{1}{2}$
where we used the fact that in worst case, the distortion is bounded
by $\frac{1}{2}$. In Section \ref{sub:The-Asymptotic-Performance},
we show that for a WSN with sufficiently large number of sensors,
the value of $\nu$ is arbitrarily close to $1$ and therefore, $\lim_{N\rightarrow\infty}D_{u}=p_{G}$.
The value of $p_{G}$ could be expressed as $p_{G}=\int_{\lambda_{t}}^{\infty}P_{b}\left(\lambda\right)f\left(\lambda|\lambda>\lambda_{t}\right)d\lambda$
where $\lambda_{t}$ is the SNR threshold applied for quantizing the
fading channel, $P_{b}\left(\lambda\right)$ is the bit error probability
for SNR of $\lambda$, and $f\left(\lambda|\lambda>\lambda_{t}\right)$
is the conditional probability distribution function of the SNR. Assuming
a binary-phase shift-keying (BPSK) modulation and an additive white
Gaussian noise with two-sided power spectral density of $\frac{N_{0}}{2}$
at the receiver, we have $P_{b}\left(\lambda\right)=Q\left(\sqrt{2\lambda}\right)$,
where $Q\left(.\right)$ represents the Q-function. From the above
results, we could bound $p_{G}$ as $p_{G}\leq Q\left(\sqrt{2\lambda_{t}}\right)$
where for obtaining this upper bound we used the fact that $P_{b}\left(\lambda\right)$
is a decreasing function of $\lambda$ and has its maximum value at
$\lambda_{t}$. In conclusion, for WSNs with sufficiently large number
of sensors, the distortion upper bound $D_{u}$ is always less than
or equal to $Q\left(\sqrt{2\lambda_{t}}\right)$. Therefore, if $\lambda_{t}$
is such that $Q\left(\sqrt{2\lambda_{t}}\right)\leq\hat{D}$, we could
conclude that our subset selection algorithm satisfies the distortion
requirement of $D\leq\hat{D}$, while reducing the average transmission
power of the sensor nodes. In this paper we assume that the condition
$p_{G}\leq\hat{D}$ holds, and proceed with presenting our subset
selection algorithm.

The rest of this paper is organized as follows. In Section \ref{sec:System-Model-and}
we present our system model used in the paper. In Section III, we
present our proposed two-phase algorithm with some examples. In Section
\ref{sec:Derivation-of-Probability}, we analytically derive the probability
distribution of the size of the minimum-size subset, as a function
of network size, channel parameters, and the source sequence length
(the size of this subset is a random variable that depends on the
CSI realizations). We also consider the computational complexity of
our analytical solution and provide suggestions to reduce this complexity
in Section V. In Section \ref{sec:Numerical-Results}, we provide
numerical results to evaluate the efficiency of our scheme in terms
of power conservation. Finally, Section \ref{sec:Conclusion} concludes
the paper.

\section{\label{sec:System-Model-and}System Model}

\begin{figure}
\centering\includegraphics[scale=0.4]{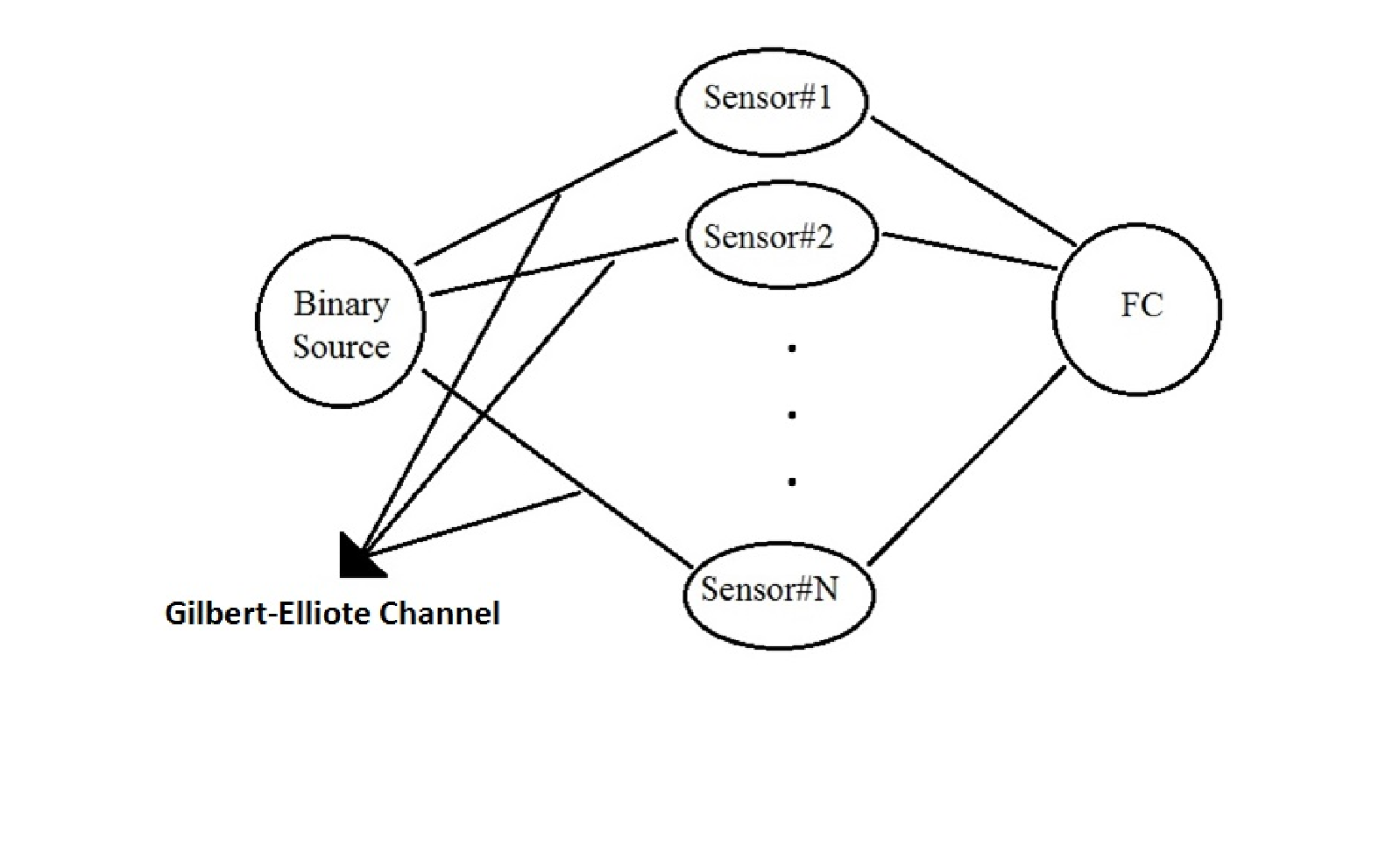}

\caption{\label{fig:System-model-of-WSN}System model of the data gathering
Wireless Sensor Network.}
\end{figure}

We consider a data gathering WSN illustrated in Fig. \ref{fig:System-model-of-WSN},
where an $M$-bit binary source is sensed by $N$ sensors via Gilbert-Elliott
channels and then transmitted to the FC via noiseless channels. To
justify the assumption of noiseless sensor-FC channels, we note that
according to IEEE 802.15.4 standard, it is recommended that the network
combines cyclic redundancy check (CRC) codes with automatic-repeat
request (ARQ) and continues re-transmission for a pre-determined number
of times \cite{Willing08}. Therefore, assuming genie CRCs, a sensor's
data is either eventually delivered to the FC error-free, or not delivered
to the FC at all. We assume that the $N$ sensors of Fig. \ref{fig:System-model-of-WSN}
are the sensors that succeeded to deliver their data to the FC before
the maximum allowed number of re-transmissions is reached. Also, note
that several researchers suggested including a forward-error correction
(FEC) scheme at sensor nodes to reduce the error probability of the
sensor-FC link (e.g., \cite{Howard06,Islam10,Abedi11,FEC-adaptive-WSN}
and references therein). This will reduce the expected number of re-transmission
requests.

\begin{figure}
\centering \includegraphics[scale=0.9]{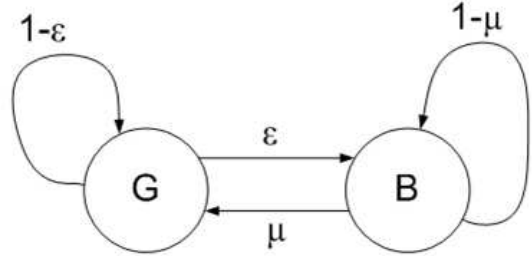} \caption{\label{fig:Gilber-Elliott-channel-model}Gilber-Elliott channel model}
\end{figure}

The state diagram of the Gilbert-Elliott channel is shown in Fig.
\ref{fig:Gilber-Elliott-channel-model}. The channel is modeled by
a Good and a Bad states and at each state the channel acts as a BSC
with transition probabilities of $p_{G}$ and $p_{B}>p_{G}$ $\left(p_{G},p_{B}<0.5\right)$,
respectively. The transition probabilities from the Good state to
the Bad state and from the Bad state to the Good state are represented
by parameters $\epsilon$ and $\mu$ respectively. As mentioned in
Section \ref{sec:Introduction}, the Gilbert-Elliott channel can be
considered as a quantized version of a correlated fading channel.
Figure \ref{fig:Realization-of-Gilbert-Elliott} shows an example
of channel realizations for a network with $N=6$ sensors and $M=256$
source bits. The dark areas show the Good state and the white areas
show the Bad state. To obtain these realizations, we generated $6$
realizations of correlated Rayleigh fading channels using Jakes model
\cite{Jakes74}. Then, we applied a quantization threshold of $\alpha_{t}=1$
on the fading amplitude, $\alpha$. The fading channels have a normalized
fading rate of $f_{d}T_{s}=2\times10^{-3}$ where $f_{d}$ is the
Doppler frequency and $T_{s}$ is the symbol period. Through Monte-Carlo
simulations, we estimated the resulting Gilbert-Elliott channel state
transition probabilities as $\epsilon=0.0075$ and $\mu=0.0041$,
respectively. It is observed from Fig. \ref{fig:Realization-of-Gilbert-Elliott}
that the CSI consists of a few runs and therefore, could be efficiently
compressed by a run-length code. In Table \ref{tab:fdTs-epsilon-mu-R}
we show the expected value of the compression rate for the run-length
coding scheme, for slowly varying fading channels with different normalized
fading rates.

\section{Proposed Two-Phase Transmission Algorithm}

Assume that we wish to re-construct the binary source at the FC with
a normalized Hamming distortion less than or equal to a threshold,
$\hat{D}$. Also assume that $p_{G}\leq\hat{D}$ and define a \textit{coverage}
event as follows:

\textbf{Definition:} \emph{A source bit is covered by a subset of
sensors if it is sensed via a Good channel by at least one of the
sensors in the subset. An $M$-bit source sequence is covered by a
subset of sensors if all of its bits are covered by the subset.}

\begin{figure}
\centering\includegraphics[scale=0.45]{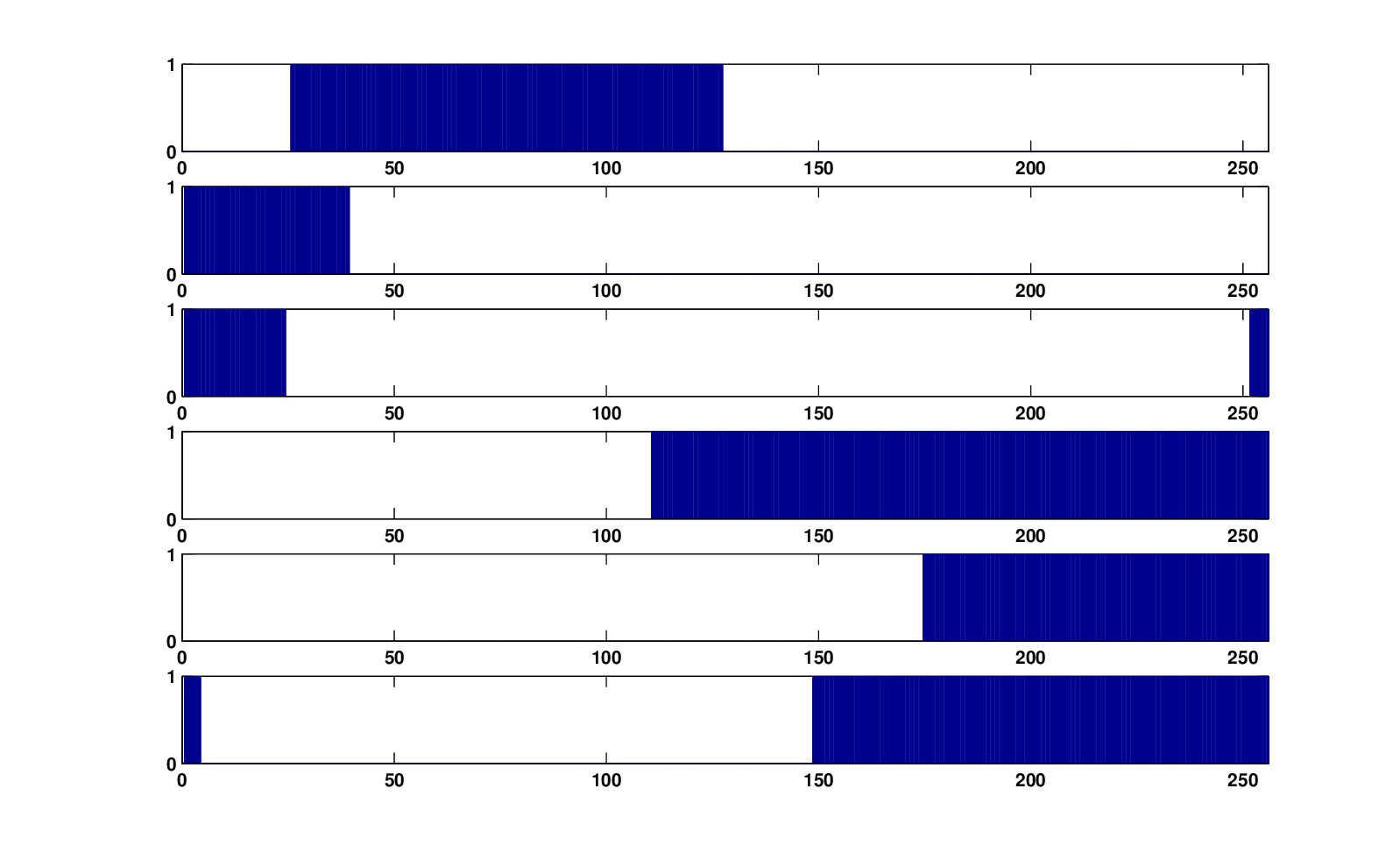}

\caption{\label{fig:Realization-of-Gilbert-Elliott}Realization of Gilbert-Elliott
Channel State Information for 6 sensors. The Gilbert-Elliott channels
are results of quantizing correlated Rayleigh Fading channels with
normalized fading rates of $f_{d}T_{s}=0.002$. A fading amplitude
of $\alpha_{t}=1$ is used as the threshold for quantization. Dark
areas show Good channel states (i.e. amplitudes above the threshold)
and white areas show Bad channel states. }
\end{figure}

\begin{figure}
\centering\includegraphics[scale=0.5]{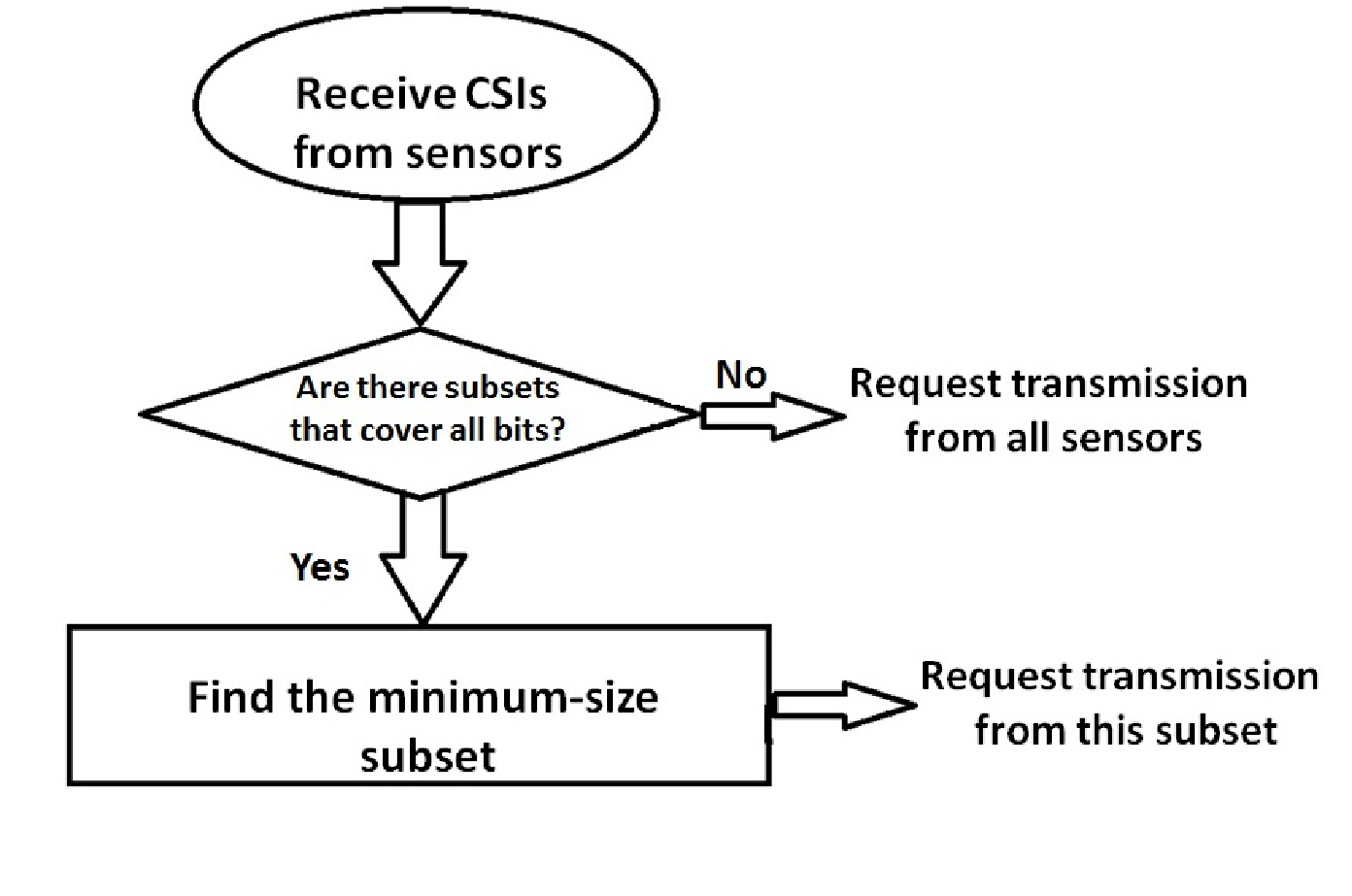}

\caption{\label{fig:The-sensor-selection}The sensor selection algorithm at
the Fusion Centre.}
\end{figure}

For example, in Fig. \ref{fig:Realization-of-Gilbert-Elliott} the
source sequence is covered by the subset consisting of the first,
second, and fourth sensors. Given the above definition, our proposed
transmission scheme is a two-phase scheme as follows. (i) At the first
phase, the sensors transmit their compressed CSI to the FC and then
wait for a feedback signal from the FC to proceed. (ii) The FC de-compresses
the received CSI and selects the smallest subset of sensors that cover
the source sequence. The implementation of the selection algorithm
at the FC is shown in Fig. \ref{fig:The-sensor-selection}. As shown
in Fig. \ref{fig:The-sensor-selection}, if no subset is covering
the source sequence, the FC requests transmission from all $N$ sensors,
in order to collect all available information for reconstructing the
source information. After selecting this minimum size subset, by transmitting
a limited feedback (e.g. an $N$-bit string where the selected sensors
are marked by $1$ and the non-selected sensors are marked by $0$)
the FC informs the sensors of which subset is selected, and only that
subset of sensors transmit their observations to the FC. It is clear
that receiving observations from this subset is sufficient to recover
the source information with a distortion less than or equal to $p_{G}$.
If $p_{G}$ is less than or equal to the tolerable distortion threshold
of the network, which we represent by $\hat{D}$, then the received
transmissions from the selected subset is sufficient to satisfy the
distortion requirement of the system. Also, by applying this subset
selection method, only a portion of sensors transmit at each time
and therefore, the average transmission power of sensors reduces.

Let us refer to the size of the selected subset by $K$. Obviously,
$K$ is a random variable that depends on the CSI realizations and
takes values from $1$ to $N$. The expected value of $K$ is an important
indicator in our proposed scheme. The ratio of this expected value
to the total number of sensors, $N$, represents the average ratio
of sensors transmitting to the FC. If this ratio becomes smaller,
the average transmission power is reduced.

To quantify the power efficiency of our proposed two-phase scheme,
we consider the total number of transmitted bits by sensors as an
indicator of the consumed power, and compare this parameter with a
\textit{conventional} one-phase scheme where all sensors transmit
all their observed bits to the FC and no CSI is transmitted. Let us
denote the total number of transmitted bits of the conventional scheme
and our scheme by $B_{1}$ and $B_{2}$, respectively. Obviously we
have $B_{1}=M\times N$. Also, it is easy to observe that $B_{2}$
is a random variable and if the expected value of the compression
rate of the run-length coding scheme is represented by $\bar{\rho}$
then, we have $E\left[B_{2}\right]=M\times\left(\bar{\rho}+E\left[K\right]\right)$.
Now, if we define an efficiency factor $\eta$ as the ratio of $B_{1}$
and $E\left[B_{2}\right]$, we have: 
\begin{equation}
\eta=\frac{N}{\bar{\rho}+E\left[K\right]}.\label{eq:eta}
\end{equation}
If $\eta$ is greater than one, then our proposed scheme consumes
less power compared to the conventional scheme. In Section \ref{sec:Numerical-Results},
we evaluate $\eta$ for Gilbert-Elliott channels with different parameters,
as well as for different number of sensors and source sequence lengths,
$M$. Our results show that in many cases, $\eta$ is considerably
larger than one.

\section{\label{sec:Derivation-of-Probability}Probability Distribution of
the Selected Subset Size}

As mentioned in Section III, the size of the selected subset is a
random variable that depends on CSI realizations. Refer to this subset
size by $K$ and let $f_{K}\left(.\right)$ and $F_{K}\left(.\right)$
be the probability mass function and the cumulative mass function
of $K$, respectively. Obviously, $F_{K}\left(k\right)$ is the probability
that there exists a subset of $k$ sensors to cover the source sequence
(if a subset of smaller size covers the sequence, we could add arbitrarily
selected sensors to this subset to make its size equal to $k$). Also,
the probability mass value, $f_{K}\left(k\right)=F_{K}\left(k\right)-F_{K}\left(k-1\right)$
is the probability that $k$ is the smallest size of a subset that
covers the source sequence. In the sequel, we derive an analytical
expression for $F_{K}\left(k\right)$.

We assume $N$ independent Gilbert-Elliott channels between source
and sensors. Let $\left(\mu_{n},\epsilon_{n}\right)$ represent the
state transition probabilities for the channel from the source to
the sensor number $n$. Assume the transmission of bit number $m$
for a fixed $m$. Let us denote the source-sensor channel states at
time interval $m$ by $\boldsymbol{C}_{m}=\left(C_{m}\left(1\right),...,C_{m}\left(N\right)\right)$
where $C_{m}\left(n\right)=1$ if the channel from the source to the
$n$th sensor is in Good state, and $C_{m}\left(n\right)=0$ if the
channel from the source to the $n$th sensor is in Bad state. Let
$S$ be an ordered subset of $(1,2,...,N)$ with cardinality $|S|$
such that $S=(S(1),S(2),...,S(|S|))$ and $S(1)<S(2)<...<S(|S|)$.
Define:

\begin{equation}
\gamma_{m}(S)=\begin{cases}
\begin{array}{c}
1;\\
0;
\end{array} & \begin{array}{c}
\prod_{m'=1}^{m}\left(\sum_{n=1}^{|S|}C_{m'}\left(S\left(n\right)\right)\right)>0\\
otherwise.
\end{array}\end{cases}\label{eq:E1}
\end{equation}
In $\left(\ref{eq:E1}\right)$, $\gamma_{m}(S)=1$ if at every bit
interval $m'=1:m$, the channel state from the source to at least
one of the sensors in set $S$ is in Good state, i.e. all $m$ bits
are covered by the set $S$. Note that if at some bit interval $m'$
all these channel states are Bad, then for that bit interval $\sum_{n=1}^{|S|}C_{m'}\left(S\left(n\right)\right)=0$
which results in $\gamma_{m}(S)=0$.

Using the above definition, $F_{K}\left(k\right)$ is equal to the
probability that there exists at least one set $S$ with $|S|\leq k$
such that $\gamma_{M}(S)=1$ ($M$ is the total number of transmitted
source bits). For calculating this probability, it is sufficient to
calculate the probability that there exists a set with $|S|=k$ and
$\gamma_{M}(S)=1$ (as mentioned above, if a set with cardinality
less than $k$ covers all bits up to bit $M$, we could add a proper
number of arbitrarily chosen sensors to make the cardinality of this
set $k$ and the extended set still covers all bits up to bit $M$).

Define $N_{k}=\left(\begin{array}{c}
N\\
k
\end{array}\right)$. There exist $N_{k}$ sets $S$ with $|S|=k$ which we refer to as
$s_{1},s_{2},...,s_{N_{k}}$. Now we can write:

\begin{equation}
F_{K}\left(k\right)=P\left(\left(\gamma_{m}\left(s_{1}\right)=1\right)||\left(\gamma_{m}\left(s_{2}\right)=1\right)||...||\left(\gamma_{m}\left(s_{N_{k}}\right)=1\right)\right)\label{eq:E2}
\end{equation}
Applying the principle of inclusion and exclusion we have:

\begin{equation}
F_{K}\left(k\right)=\sum_{i}P\left(\gamma_{m}\left(s_{i}\right)=1\right)-\sum_{\begin{array}{c}
i,j\\
i\neq j
\end{array}}P\left(\left(\gamma_{m}\left(s_{i}\right)=1\right),\left(\gamma_{m}\left(s_{j}\right)=1\right)\right)+...\label{eq:E3}
\end{equation}

To simplify the notation, let us define $W=\left(W(1),W(2),...,W(|W|)\right)$
as an ordered subset of $\left(1,2,...,N_{k}\right)$, where $\left|W\right|\leq N_{k}$
is the cardinality of $W$. Now consider the sets $s_{W(1)},s_{W(2)},...,s_{W(|W|)}$.
It is clear that 
\[
P\left(\left(\gamma_{m}\left(s_{W(1)}\right)=1\right),\left(\gamma_{m}\left(s_{W(2)}\right)=1\right),...,\left(\gamma_{m}\left(s_{W(|W|)}\right)=1\right)\right)=P\left(\prod_{l=1}^{|W|}\gamma_{m}\left(s_{W\left(l\right)}\right)=1\right).
\]
Let us also define:

\begin{equation}
\Gamma_{m}\left(W\right)=\prod_{l=1}^{|W|}\gamma_{m}\left(s_{W\left(l\right)}\right).\label{eq:E4}
\end{equation}
Now, by noting that there are $2^{N_{k}}$ possible choices for $W$,
which we represent by $w_{1},w_{2},...,w_{2^{N_{k}}}$, one can rewrite
the inclusion-exclusion expression of (\ref{eq:E3}) as follows:

\begin{equation}
F_{K}\left(k\right)=\sum_{j=1}^{2^{N_{k}}}\left(-1\right)^{|w_{j}|+1}Pr\left(\Gamma_{m}\left(w_{j}\right)=1\right).\label{eq:E5}
\end{equation}

Now let us look at vector $\boldsymbol{C}_{m}$ defined above. There
are $2^{N}$ possible realizations for $\boldsymbol{C}_{m}$ which
are in fact the $2^{N}$ distinct binary $n$-tuples. We refer to
these binary $n$-tuples by $\boldsymbol{u}_{1},\boldsymbol{u}_{2},...,\boldsymbol{u}_{2^{N}}$.
Now the joint probability of events $\Gamma_{m}\left(w_{j}\right)=1$
and $\boldsymbol{C}_{m}=\boldsymbol{u}_{i}$ can be calculated as:
\begin{equation}
P\left(\Gamma_{m}\left(w_{j}\right)=1,\boldsymbol{C}_{m}=\boldsymbol{u}_{i}\right)=\sum_{l=1}^{2^{N}}P\left(\Gamma_{m}\left(w_{j}\right)=1,\boldsymbol{C}_{m}=\boldsymbol{u}_{i},\boldsymbol{C}_{m-1}=\boldsymbol{u}_{l}\right)\label{eq:P_joint_gamma_u_1}
\end{equation}
where we can write: 
\begin{equation}
P\left(\Gamma_{m}\left(w_{j}\right)=1,\boldsymbol{C}_{m}=\boldsymbol{u}_{i},\boldsymbol{C}_{m-1}=\boldsymbol{u}_{l}\right)=P\left(\Gamma_{m}\left(w_{j}\right)=1,\Gamma_{m-1}\left(w_{j}\right)=1,\boldsymbol{C}_{m}=\boldsymbol{u}_{i},\boldsymbol{C}_{m-1}=\boldsymbol{u}_{l}\right).\label{eq:P_joint_gamma_u_2}
\end{equation}
Note that in (\ref{eq:P_joint_gamma_u_2}) 
\[
\begin{array}{c}
P\left(\Gamma_{m}\left(w_{j}\right)=1,\Gamma_{m-1}\left(w_{j}\right)=1,\boldsymbol{C}_{m}=\boldsymbol{u}_{i},\boldsymbol{C}_{m-1}=\boldsymbol{u}_{l}\right)\\
=P\left(\Gamma_{m}\left(w_{j}\right)=1,\boldsymbol{C}=\boldsymbol{u}_{i},\boldsymbol{C}_{m-1}=\boldsymbol{u}_{l}\right)\times P\left(\Gamma_{m-1}\left(w_{j}\right)=1,|\Gamma_{m}\left(w_{j}\right)=1,\boldsymbol{C}_{m}=\boldsymbol{u}_{i},\boldsymbol{C}_{m-1}=\boldsymbol{u}_{l}\right)
\end{array}
\]
and $P\left(\Gamma_{m-1}\left(w_{j}\right)=1|\Gamma_{m}\left(w_{j}\right)=1\right)=1$.

Using (\ref{eq:P_joint_gamma_u_2}), one can obtain 
\begin{equation}
\begin{array}{cc}
P\left(\Gamma_{m}\left(w_{j}\right)=1,\boldsymbol{C}_{m}=\boldsymbol{u}_{i},\boldsymbol{C}_{m-1}=\boldsymbol{u}_{l}\right)=\\
P\left(\Gamma_{m-1}\left(w_{j}\right)=1,\boldsymbol{C}_{m-1}=\boldsymbol{u}_{l}\right)\times P\left(\Gamma_{m}\left(w_{j}\right)=1,\boldsymbol{C}_{m}=\boldsymbol{u}_{i}|\Gamma_{m-1}\left(w_{j}\right)=1,\boldsymbol{C}_{m-1}=\boldsymbol{u}_{l}\right).
\end{array}\label{eq:P_joint_gamma_u_3}
\end{equation}
Let us rewrite the second term in the righthand side of (\ref{eq:P_joint_gamma_u_3})
as follows: 
\begin{equation}
\begin{array}{ccc}
P\left(\Gamma_{m}\left(w_{j}\right)=1,\boldsymbol{C}_{m}=\boldsymbol{u}_{i}|\Gamma_{m-1}\left(w_{j}\right)=1,\boldsymbol{C}_{m-1}=\boldsymbol{u}_{l}\right)=\\
P\left(\boldsymbol{C}_{m}=\boldsymbol{u}_{i}|\Gamma_{m-1}\left(w_{j}\right)=1,\boldsymbol{C}_{m-1}=\boldsymbol{u}_{l}\right)\times P\left(\Gamma_{m}\left(w_{j}\right)=1|\boldsymbol{C}_{m}=\boldsymbol{u}_{i},\Gamma_{m-1}\left(w_{j}\right)=1,\boldsymbol{C}_{m-1}=\boldsymbol{u}_{l}\right).
\end{array}\label{eq:P_joint_gamma_u_4}
\end{equation}
Note in the righthand side of (\ref{eq:P_joint_gamma_u_4}) that given
$\boldsymbol{C}_{m-1}$, $\boldsymbol{C}_{m}$ is independent of $\Gamma_{m-1}\left(w_{j}\right)$.
Also given $\Gamma_{m-1}\left(w_{j}\right)$ and $\boldsymbol{C}_{m}$,
$\Gamma_{m}\left(w_{j}\right)$ is independent of $\boldsymbol{C}_{m-1}$.
This second claim is made by noting that if $\Gamma_{m-1}\left(w_{j}\right)=1$,
then $\Gamma_{m}\left(w_{j}\right)=1$ if and only if given the channel
realization $\boldsymbol{C}_{m}$, $w_{j}$ is such that for every
subset $s_{w_{j}\left(i\right)}$, $i=1:|w_{j}|$, at least one of
the sensors in the subset has a Good source-sensor channel. Therefore,
it is clear that $P\left(\Gamma_{m}\left(w_{j}\right)=1|\boldsymbol{C}_{m}=\boldsymbol{u}_{i},\Gamma_{m-1}\left(w_{j}\right)=1,\boldsymbol{C}_{m-1}=\boldsymbol{u}_{l}\right)$
is a function of channel realization $\boldsymbol{u}_{i}$ and the
set $w_{j}$. If we refer to this function by $d_{j}\left(i\right)$,
one can write:

\begin{equation}
d_{j}\left(i\right)=\begin{cases}
\begin{array}{c}
1;\\
0;
\end{array} & \begin{array}{c}
\prod_{l=1}^{|w_{j}|}\left(\sum_{h=1}^{k}\boldsymbol{u}_{i}\left(s_{w_{j}\left(l\right)}\left(h\right)\right)\right)>0\\
otherwise
\end{array}\end{cases}\label{eq:E7}
\end{equation}
To clarify this definition, note that if $\sum_{h=1}^{k}\boldsymbol{u}_{i}\left(s_{w_{j}\left(l\right)}\left(h\right)\right)$
is a positive number, then given $\boldsymbol{C}_{m}=\boldsymbol{u}_{i}$,
the subset $s_{w_{j}\left(l\right)}$ covers the $m$th bit. In fact
(\ref{eq:E7}) states that $d_{j}\left(i\right)$ is $1$ if for every
set $s_{w_{j}(l)}$, at least one of the sensors in this set receives
the $m$th bit through a Good channel.

Let us define a matrix $Q=\left[q\left(i,l\right)\right]$ where $q\left(i,l\right)=P\left(\boldsymbol{C}_{m}=\boldsymbol{u}_{i}|\boldsymbol{C}_{m-1}=\boldsymbol{u}_{l}\right)$.
From the channel model, we can observe that:

\begin{equation}
q\left(i,l\right)=\prod_{n=1}^{N}P\left(C_{m}\left(n\right)=u_{i}\left(n\right)|C_{m-1}\left(n\right)=u_{l}\left(n\right)\right)\label{eq:E8}
\end{equation}
and $P\left(C_{m}\left(n\right)=u_{i}\left(n\right)|C_{m-1}\left(n\right)=u_{l}\left(n\right)\right)$
is readily expressed based on the $n$th source-sensor channel state
transition probabilities $\left(\mu_{n},\epsilon_{n}\right)$.

Now if we define a matrix 
\[
A_{j}=\left[a_{j}\left(i,l\right)\right],
\]
where $a_{j}\left(i,l\right)=d_{j}\left(i\right)q\left(i,l\right)$,
using (\ref{eq:P_joint_gamma_u_4}) and above discussion, one can
note that 
\begin{equation}
P\left(\Gamma_{m}\left(w_{j}\right)=1,\boldsymbol{C}_{m}=\boldsymbol{u}_{i}|\Gamma_{m-1}\left(w_{j}\right)=1,\boldsymbol{C}_{m-1}=\boldsymbol{u}_{l}\right)=a_{j}\left(i,l\right)
\end{equation}
and hence (\ref{eq:P_joint_gamma_u_3}) can be represented as:

\begin{equation}
P\left(\Gamma_{m}\left(w_{j}\right)=1,\boldsymbol{C}_{m}=\boldsymbol{u}_{i},\boldsymbol{C}_{m-1}=\boldsymbol{u}_{l}\right)=P\left(\Gamma_{m-1}\left(w_{j}\right)=1,\boldsymbol{C}_{m-1}=\boldsymbol{u}_{l}\right)\times a_{j}\left(i,l\right).\label{eq:P_joint_gamma_u_5}
\end{equation}

To simplify (\ref{eq:P_joint_gamma_u_5}), let us define a vector
\[
\boldsymbol{X}_{m}=(X_{m}\left(1\right),X_{m}\left(2\right),...,X_{m}\left(2^{N}\right))
\]
where 
\[
X_{m}\left(i\right)=Pr\left(\Gamma_{m}\left(w_{j}\right)=1,\boldsymbol{C}_{m}=\boldsymbol{u}_{i}\right).
\]
Now from (\ref{eq:P_joint_gamma_u_1}) and (\ref{eq:P_joint_gamma_u_5})
we have: 
\begin{equation}
X_{m}\left(i\right)=\sum_{l=1}^{2^{N}}a_{j}\left(i,l\right)X_{m-1}(l)\label{eq:E9}
\end{equation}
which leads to the following recursive matrix equation: 
\begin{equation}
\boldsymbol{X}_{m}=A_{j}\boldsymbol{X}_{m-1}.\label{eq:E10}
\end{equation}
Note that to simplify the notation, we dropped dependence of $\boldsymbol{X}_{m}$
to $j$. Also note that $A_{j}$ is constructed by forcing some rows
of matrix $Q$ to zero. Those are the rows $i$ such that $d_{j}(i)=0$.
Now from (\ref{eq:E10}), we arrive at the following solution for
$X_{m}$:

\begin{equation}
\boldsymbol{X}_{m}=A_{j}^{m-1}\boldsymbol{X}_{1}\label{eq:E11}
\end{equation}
where $A_{j}^{m-1}$ is the $m-1$ power of matrix $A_{j}$, and the
initial vector $\boldsymbol{X}_{1}$ is expressed as:

\begin{equation}
X_{1}\left(i\right)=P\left(\Gamma_{1}\left(w_{j}\right)=1,\boldsymbol{C}_{1}=\boldsymbol{u}_{i}\right)=P\left(\boldsymbol{C}_{1}=\boldsymbol{u}_{i}\right)P\left(\Gamma_{1}\left(w_{j}\right)=1|\boldsymbol{C}_{1}=\boldsymbol{u}_{i}.\right)\label{eq:E12}
\end{equation}
It is straightforward to show that 
\begin{equation}
P\left(\Gamma_{1}\left(w_{j}\right)=1|\boldsymbol{C}_{1}=\boldsymbol{u}_{i}\right)=d_{j}\left(i\right).
\end{equation}
Now noting the independence assumption for source-sensor channels,
we have: 
\begin{equation}
X_{1}\left(i\right)=d_{j}\left(i\right)\times\prod_{n=1}^{N}P\left(C_{1}\left(n\right)=u_{i}\left(n\right)\right)\label{eq:X1_i_continued}
\end{equation}
Following \cite{Sadeghi_Fading_Review} we let the initial channel
state $C_{1}\left(n\right)$ have the steady state probability distribution
of the corresponding Markov process. For the Markov process of Fig.
\ref{fig:Gilber-Elliott-channel-model} this steady state distribution
is as follows: 
\begin{equation}
P\left(C_{1}\left(n\right)=1\right)=\frac{\mu_{n}}{\epsilon_{n}+\mu_{n}}
\end{equation}
and 
\begin{equation}
P\left(C_{1}\left(n\right)=0\right)=\frac{\epsilon_{n}}{\epsilon_{n}+\mu_{n}}
\end{equation}

After solving (\ref{eq:E12}), we calculate $\Gamma_{M}\left(w_{j}\right)$
as follows: 
\begin{equation}
\Gamma_{M}\left(w_{j}\right)=\sum_{i=1}^{2^{N}}X_{M}\left(i\right)
\end{equation}
and by substituting in (\ref{eq:E5}), we can evaluate $F_{K}\left(k\right)$.

\begin{figure}
\centering\includegraphics[scale=0.6]{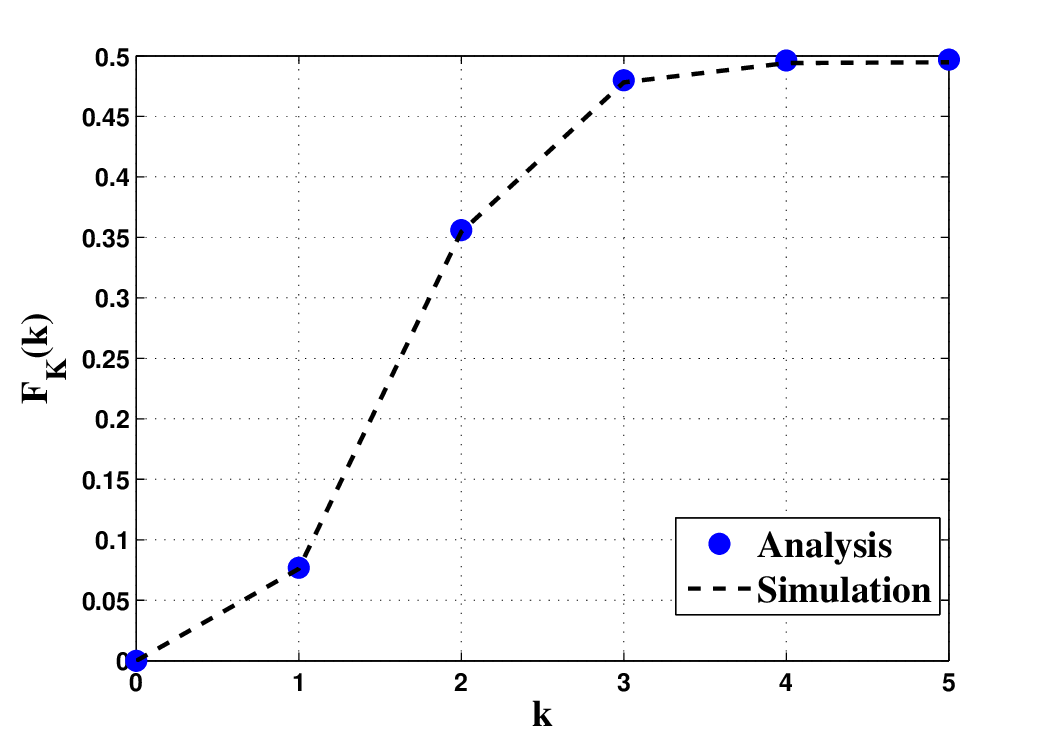}

\caption{\label{fig:cmf-comparisons-analysis-sim}The cumulative mass function
of selected subset size, $K$, for a network with $N=5$ sensors and
source sequence length of $M=128$ bits. The source-sensor channels
have identical transition probabilities of $\left(\mu_{n},\epsilon_{n}\right)=\left(0.0191,0.0256\right)$.}
\end{figure}

To assess the accuracy of our analyses, in Fig. \ref{fig:cmf-comparisons-analysis-sim}
we compare $F_{K}\left(k\right)$ found using (\ref{eq:E5}) with
simulations. For these simulations, $10^{5}$ source sequences of
length $M=128$ bits are transmitted to $N=5$ sensors via identically
distributed Gilbert-Elliott channels with parameters $\left(\mu_{n},\epsilon_{n}\right)=\left(0.0191,0.0256\right)$
and $10^{5}$ realizations of $K$ are generated by comparing the
$5$ corresponding CSIs. The channel parameters $\left(\mu_{n},\epsilon_{n}\right)$
are taken from Table \ref{tab:fdTs-epsilon-mu-R} (see Section \ref{sec:Numerical-Results}).
It is clear from these results that our analysis is in excellent agreement
with the simulated results.

As observed from Fig. \ref{fig:cmf-comparisons-analysis-sim}, $F_{K}\left(5\right)\simeq0.5$,
i.e., in almost $50\%$ of the time, employing all five sensors is
not sufficient to cover all source bits. However, as shown in Fig.
\ref{fig:The-sensor-selection}, in these cases, our algorithm forces
all sensors to transmit their observations to the FC, i.e. we force
$F_{K}\left(N\right)=1$. In the following section, we will show that
by increasing $N$, the coverage probability increases where the actual
values of $F_{K}\left(N\right)$ (before forcing to one) are much
closer to one.

Now, the expected value of $K$ can be expressed as:

\[
E\left[K\right]=\sum_{k=1}^{N}k\times f_{K}\left(k\right)=N\times F_{K}\left(N\right)-\sum_{k=1}^{N-1}F_{K}\left(k\right)
\]
where by noting $F_{K}\left(N\right)=1$ for our scheme, we reach:

\begin{equation}
E\left[K\right]=N-\sum_{k=1}^{N-1}F_{K}\left(k\right).\label{eq:EK_based_on_FK}
\end{equation}

In Section \ref{sec:Numerical-Results}, we use the expected value
of the subset size, $E\left[K\right]$, to evaluate the power reduction
achieved by our proposed algorithm.

\section{Complexity and Asymptotic Performance of the Proposed Algorithm }

In what follows, we analyze the computational complexity and asymptotic
performance of the proposed two-phase transmission algorithm.

\subsection{Complexity Considerations}

Calculating $F_{K}\left(k\right)$ from (\ref{eq:E5}) introduces
a computational complexity that is exponentially increasing by $N_{k}$
where $N_{k}=\left(\begin{array}{c}
N\\
k
\end{array}\right)$%
\footnote{Note that this computational complexity only applies to our analysis.
Implementing the algorithm at the FC is considerably less complex
as in that case the FC has the CSI realizations and only needs to
compare them to find the minimum size subset.%
}. Calculation of $F_{K}\left(k\right)$ and consequently, $E\left[K\right]$
is time-consuming for large values of $N$. In fact, the run time
for networks with more than $N=7$ sensors is very large. Therefore,
it is desired to introduce bounds on $E\left[K\right]$. It is possible
to introduce two simple upper bounds on $E\left[K\right]$ as follows:

Let $\tilde{F_{K}}\left(k\right)$ be a lower bound for $F_{K}\left(k\right)$.
Then, from (\ref{eq:EK_based_on_FK}) one can find an upper bound
as follows:

\begin{equation}
E\left[K\right]\leq N-\sum_{k=1}^{N-1}\tilde{F_{K}}\left(k\right).\label{eq:Ek_upper_bound_1}
\end{equation}
One possible choice for $\tilde{F_{K}}\left(k\right)$ is by applying
Bonferroni's lower bound \cite{Bonferroni36}. Let $L_{k}\leq N_{k}/2$
be an integer, then the inclusion-exclusion formula of (\ref{eq:E5})
can be lower-bounded as $F_{K}\left(k\right)\geq\tilde{F_{K}}\left(k\right)$
where

\begin{equation}
\tilde{F_{K}}\left(k\right)=\sum_{\begin{array}{c}
j=1\\
|w_{j}|\leq2L_{k}
\end{array}}^{2^{N_{k}}}\left(-1\right)^{|w_{j}|+1}Pr\left(\Gamma_{m}\left(w_{j}\right)=1\right).\label{eq:F_tild_Bonferroni}
\end{equation}
Through simulations, we concluded that for values of $L_{k}$ which
introduce a reasonable computational complexity, the bound of (\ref{eq:F_tild_Bonferroni})
is not tight and in fact leads to a negative value in most cases.

Another simple upper bound can be derived by noting that $F_{K}\left(k\right)\geq F_{K}\left(1\right)$,
for $k=1:N-1$, which by using (\ref{eq:EK_based_on_FK}) leads to:

\begin{equation}
E\left(K\right)\leq N-\left(N-1\right)F_{K}\left(1\right)\label{eq:Ek_Fk1}
\end{equation}
where $F_{K}\left(1\right)$ is the probability that one sensor covers
the source sequence (i.e., the probability that at least one of the
$N$ sensors receives all $M$ source bits via Good source-sensor
channels). Fortunately the value of $F_{K}\left(1\right)$ can be
simply derived as follows. The probability that the $n$th sensor
covers all source bits equals the probability that the corresponding
source-sensor channel is initially at a Good state and stays at this
state for the next $M-1$ bit intervals. This probability is equal
$\left(\frac{\mu_{n}}{\mu_{n}+\epsilon_{n}}\right)\left(1-\epsilon_{n}\right)^{M-1}$.
Therefore, the probability that none of the $N$ sensors covers the
source sequence equal $\prod_{n=1}^{N}\left(1-\left(\frac{\mu_{n}}{\mu_{n}+\epsilon_{n}}\right)\left(1-\epsilon_{n}\right)^{M-1}\right)$
and eventually, the probability that at least one of these $N$ sensors
covers the source sequence is given by:

\begin{equation}
F_{K}\left(1\right)=1-\prod_{n=1}^{N}\left(1-\left(\frac{\mu_{n}}{\mu_{n}+\epsilon_{n}}\right)\left(1-\epsilon_{n}\right)^{M-1}\right).\label{eq:FK_1}
\end{equation}
Note that if all source-sensor channels have identical parameters
$\left(\mu_{n},\epsilon_{n}\right)=\left(\mu,\epsilon\right)$, it
is easy to verify that $F_{K}\left(1\right)$ is a monotonically increasing
function of $N$. This is expected, as by increasing the number of
sensors, there is a higher probability that at least one of these
sensors covers the source sequence.

By replacing $F_{K}\left(1\right)$ from (\ref{eq:FK_1}) in (\ref{eq:Ek_Fk1}),
we find a simple upper bound for $E\left[K\right]$ as follows,

\begin{equation}
E\left(K\right)\leq N-\left(N-1\right)\left(1-\prod_{n=1}^{N}\left(1-\left(\frac{\mu_{n}}{\mu_{n}+\epsilon_{n}}\right)\left(1-\epsilon_{n}\right)^{M-1}\right)\right).\label{eq:Ek_upper_bound}
\end{equation}

\begin{figure}
\centering\includegraphics[scale=0.7]{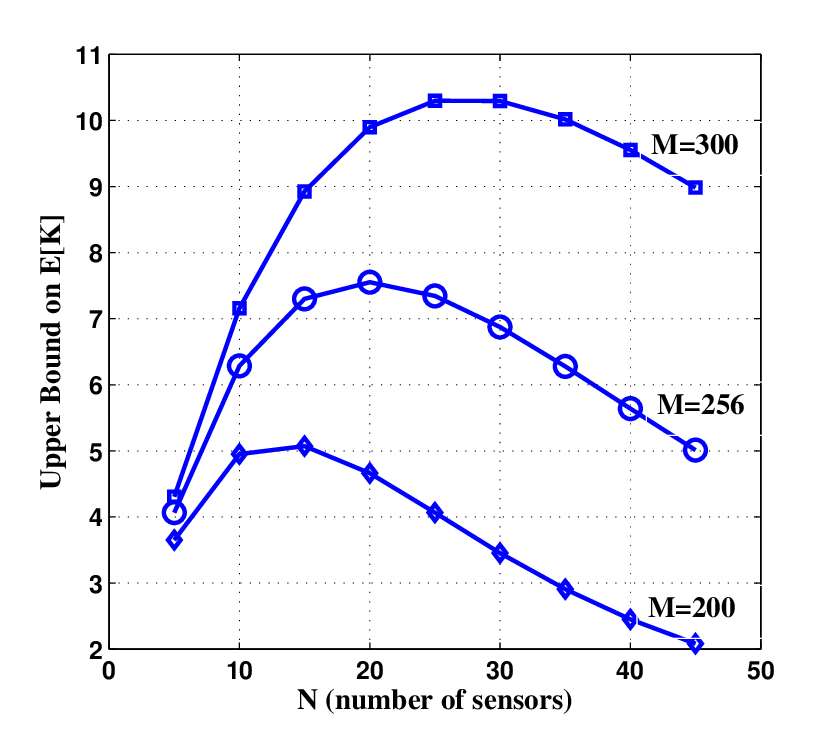}

\caption{\label{fig:Ek_upperbound}The upper bound on $E\left[K\right]$ for
a network with identical source-sensor channel parameters $\left(\mu,\epsilon\right)=\left(0.0041,0.0075\right)$.
The source sequence lengths are $M=200$, $M=256$, and $M=300$ bits.}
\end{figure}

Figure \ref{fig:Ek_upperbound} shows $E\left[K\right]$ as a function
of $N$ for a network with identical source-sensor channel parameters
$\left(\mu,\epsilon\right)=\left(0.0041,0.0075\right)$ and source
sequence lengths of $M=200,\,256,\,300$ bits. The values of $\left(\mu,\epsilon\right)$
are based on Table \ref{tab:fdTs-epsilon-mu-R}. Note the non-monotonic
behaviour that is observed in Fig. \ref{fig:Ek_upperbound} for the
upper bound of $E\left[K\right]$. This non-monotonic behaviour is
due to two reasons. The first is based on the fact that this upper
bound is not tight. However, there is another rational behind the
non-monotonic behaviour of this upper bound. That is, in cases where
the source is not covered by any subset, we demand transmission from
all $N$ sensors (i.e., $K=N$). When $N$ increases, there are subsets
with larger sizes to examine for possible coverage. Therefore, $E\left[K\right]$
might increase in such cases. Although the upper bound of (\ref{eq:Ek_upper_bound})
is not tight, as we will see in Section \ref{sec:Numerical-Results},
even by applying this simple bound, we observe considerable power
reduction when employing our proposed algorithm for networks with
large values of $N$.

\subsection{\label{sub:The-Asymptotic-Performance}Asymptotic Performance}

Here, we consider the asymptotic performance of our proposed algorithm
for large values of $N$. For simplicity, let us assume that all $N$
source-sensor channels have identical state transition probabilities
$\left(\mu,\epsilon\right)$. It is clear from (\ref{eq:FK_1}) that
for identical values of $\left(\mu_{n},\epsilon_{n}\right)=\left(\mu,\epsilon\right)$,
$F_{K}\left(1\right)$ is a monotonically increasing function of $N$
and $\lim_{n\rightarrow\infty}F_{K}\left(1\right)=1$. By noting that
$F_{K}\left(1\right)\leq F_{K}\left(k\right)\leq1$ for $k=2:N$,
the lower bounds of $\tilde{F}_{K}\left(k\right)=F_{K}\left(1\right),\, k=2:N$
are asymptotically tight. Therefore, the upper bound of (\ref{eq:Ek_upper_bound})
is asymptotically tight. If we let $\left(\mu_{n},\epsilon_{n}\right)=\left(\mu,\epsilon\right)$
and by taking the derivative of (\ref{eq:Ek_upper_bound}) with respect
to $N$, one can find a value $N_{0}$ such that for all $N\geq N_{0}$,
the upper bound of $E\left[K\right]$ is monotonically decreasing
by $N$ 
\begin{equation}
N_{0}=\left\lceil 1+\frac{1}{\ln\frac{1}{1-x}}\right\rceil 
\end{equation}
where $x=\left(\frac{\mu}{\mu+\epsilon}\right)\left(1-\epsilon\right)^{M-1}$.
From the above discussion, we conclude that for sufficiently large
values of $N$, $\frac{E\left[K\right]}{N}$ is a monotonically decreasing
function of $N$ (decaying by a rate of $\frac{1}{N}$ or faster).
Rewrite (\ref{eq:eta}) as 
\begin{equation}
\frac{1}{\eta}\leq\frac{\bar{\rho}}{N}+\frac{E\left[K\right]}{N}
\end{equation}
and note that $\bar{\rho}\leq1$. We observe that $\frac{1}{\eta}$,
which is the ratio of power consumption for our proposed algorithm
to the conventional transmission scheme, decays by increasing $N$
(at least by a rate of $\frac{1}{N}$). Therefore, our proposed algorithm
becomes asymptotically more power efficient by increasing $N$.

At the end of this section, we note that when we were motivating the
idea in Section \ref{sec:Introduction}, we applied a parameter $\nu$
for bounding the distortion, where we defined $\nu$ as the probability
that there exists a (minimum-size) subset that covers all source bits.
We claimed that for sufficiently large $N$, $\nu$ can be arbitrarily
close to one. To prove this claim, note that the probability that
such subset exists, is greater than or equal the probability that
such subset exists and its size is $k$ (for an arbitrarily chosen
$k\leq N$). Therefore, $\nu\geq F_{K}\left(k\right)\geq F_{K}\left(1\right)$
and $F_{K}\left(1\right)$ could be arbitrarily close to one, given
a sufficiently large $N$.

\section{\label{sec:Numerical-Results}Numerical Results}

In this section, we provide some numerical results to evaluate the
power efficiency of our proposed algorithm. In this work, and without
loss of generality, we only consider cases where all source-sensor
channels have identical parameters $\left(\mu,\epsilon\right)$. The
Gilbert-Elliott channel parameters are derived by simulating a correlated
Rayleigh fading channel using Jakes model and then quantizing the
simulated channel by assuming a threshold on the fading amplitude.
If we represent the fading amplitude by $\alpha$ and assume that
the source is transmitting each bit with energy $E_{b}$ and the AWGN
has a one-sided power spectral density of $N_{0}$, then the instantaneous
received SNR equals$\frac{\alpha^{2}E_{b}}{N_{0}}$ at the sensor.
We consider a threshold of $\alpha_{thr}=1$. That is we assume that
SNRs above $\frac{E_{b}}{N_{0}}$ leads to a Good delivery of the
source bit to the sensor (i.e, the probability of detection error,
$p_{G}$, is sufficiently low to have $p_{G}\leq\hat{D}$ as discussed
in Section \ref{sec:Introduction}). The assumption of $\alpha_{thr}=1$
is justified as follows. If we assume that the channel phase shift
is perfectly estimated and compensated at the sensor node, then for
all $\alpha>\alpha_{thr}$, the channel provides error detection probabilities
less than or equal to the error detection probability of an AWGN channel
with SNR of $\frac{E_{b}}{N_{0}}$. Therefore, by setting this threshold,
we eliminate the non-constructive effect of fading and provided source-sensor
channels with link qualities equivalent or superior to an AWGN channel.
We consider a slow-fading channel, i.e., channels with the normalized
fading rates of $f_{d}T_{s}\leq0.01$, where $f_{d}$ is the maximum
Doppler shift and $T_{s}$ is the symbol duration. The reason we consider
slow-fading channels is that as discussed in previous sections, for
these channels the run-length coding of CSI sequences provides an
efficient compression.

As discussed earlier, the parameters $\left(\mu,\epsilon\right)$
for the Gilbert-Elliott channel are estimated using Monte-Carlo simulation
of sufficiently large number of realizations of the fading channel
amplitude. For values of $f_{d}T_{s}=0.002,0.005,0.008$, the corresponding
values of $\left(\mu,\epsilon\right)$ are shown in Table \ref{tab:fdTs-epsilon-mu-R}.
Table \ref{tab:fdTs-epsilon-mu-R} also shows the expected value of
compression rate for CSIs, for different sequence lengths of $M=128,\, M=256$.
As expected, the compression rate decreases when increasing $M$.

Tables \ref{tab:EK_eta_M_128} and \ref{tab:EK_eta_M_256} show values
of $\left(E\left[K\right],\eta\right)$ for networks with $N=4,5,6$
sensors. We observe that $E\left[K\right]$ is a non-monotonic function
of $N$. The justification of this non-monotonic behaviour was discussed
in Section \ref{sub:The-Asymptotic-Performance} and Fig. \ref{fig:Ek_upperbound}.
Note that the efficiency factor, $\eta$, is monotonically increasing
function of $N$, which confirms the increase in efficiency of our
proposed algorithm as the number of sensors, $N$, increases.

From Tables \ref{tab:EK_eta_M_128} and \ref{tab:EK_eta_M_256}, it
is clear that our algorithm is more efficient for channels with slower
fading rates. For instance, in Table \ref{tab:EK_eta_M_128}, if we
let $N=5$, we observe that the efficiency factor for channels with
$f_{d}T_{s}=0.002$ is $1.92$ which shows an almost two-fold decrease
in power consumption achieved by our algorithm compared to the conventional
transmission scheme. However, when we increase $f_{d}T_{s}$ to $0.008$,
$\eta$ decreases to $1.30$. Also, by comparing the results of Table
\ref{tab:EK_eta_M_128} and Table \ref{tab:EK_eta_M_256}, we observe
that our proposed algorithm is more efficient for shorter source sequence
lengths, $M$. The reason is that we defined a coverage event as the
event that all source bits are covered. Therefore, the coverage probability
of a subset decreases by increasing $M$. This results in an increase
of $E\left[K\right]$ and consequently a decrease in $\eta$. The
only case where our scheme shows an inferior performance to the conventional
transmission scheme is for $M=256$, $f_{d}T_{s}=0.008$, and $N=4$,
where $\eta=0.99$.

\begin{figure}
\centering\includegraphics[scale=0.7]{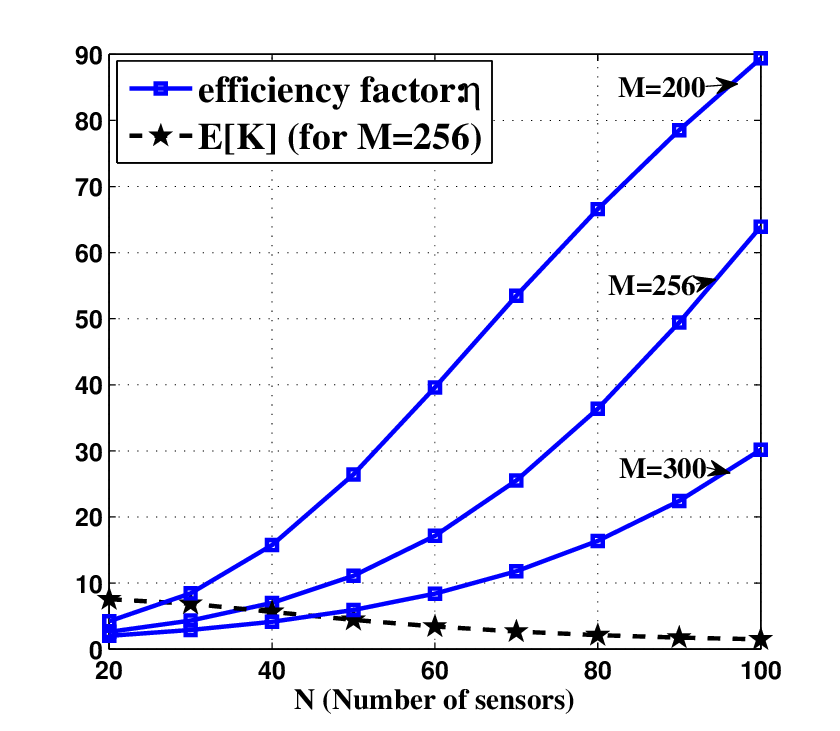}

\caption{\label{fig:Ek_eta_large_N}Values of $\eta$ for networks with source
sequence lengths of$M=200,\,256,\,300$ bits and $f_{d}T_{s}=0.002$.
The dashed line shows $E\left[K\right]$ for $M=256$ bits.}
\end{figure}

To examine the case of large networks with large values of $N$, we
turn to the upper bound of (\ref{eq:Ek_upper_bound}). Replacing this
upper bound in (\ref{eq:eta}) provides a lower bound on $\eta$.
Figure \ref{fig:Ek_eta_large_N} shows the upper bound of $E\left[K\right]$
and the resulting lower bound on $\eta$ for networks with source
sequence length of $M=256$ bits and $f_{d}T_{s}=0.002$. One can
note the considerable gains for these large values of $N$ when using
our algorithm. For example, for a network with $N=50$ sensors, our
proposed algorithm provides at least a twelve-fold decrease in the
consumed power compared to the conventional transmission scheme with
all nodes transmitting ($\eta>12$). To examine the effect of different
block lengths on $\eta$, we also consider block lengths $M=200$
and $M=300$ in Fig. \ref{fig:Ek_eta_large_N}. As observed, the efficiency
factor decreases by increasing the block length. This is due the fact
that as $M$ increases, the probability that $k$ sensors cover all
$M$ bits decreases. As a result $E\left[K\right]$ increases and
$\eta$ becomes smaller. Nonetheless, we observe that for $M=300$
and $N=50$, our algorithm has an efficiency factor close to $6$.

\begin{table}
\caption{\label{tab:fdTs-epsilon-mu-R} Gilbert-Elliott channel transition
probabilities and achieved compression rates by run-length coding
scheme for different values of the normalized fading rate, $f_{d}T_{s}$.
The fading amplitude threshold for deciding between Good and Bad states
is set to $1$. The source sequence lengths of $M=128$ and $M=256$
bits are considered.}

\centering%
\begin{tabular}{|c|c|c|c|c|}
\hline 
$f_{d}T_{s}$  & $\epsilon$  & $\mu$  & $\bar{\rho}(M=128)$  & $\bar{\rho}(M=256)$\tabularnewline
\hline 
\hline 
$0.002$  & $0.0075$  & $0.0041$  & $0.1071$  & $0.0813$\tabularnewline
\hline 
$0.005$  & $0.0165$  & $0.0112$  & $0.1630$  & $0.1454$\tabularnewline
\hline 
$0.008$  & $0.0256$  & $0.0191$  & $0.2223$  & $0.2134$\tabularnewline
\hline 
\end{tabular}
\end{table}

\begin{table}
\caption{\label{tab:EK_eta_M_128}Values of$\left(E\left[K\right],\eta\right)$
for networks with $N=4,5,6$ sensors and source sequence length of$M=128$
bits.}

\centering%
\begin{tabular}{|c||c|c|c|}
\hline 
\multicolumn{1}{|c||||||}{$f_{d}T_{s}$} & $N=4$  & $N=5$  & $N=6$\tabularnewline
\hline 
$0.002$  & $\left(2.44,1.57\right)$  & $\left(2.49,1.92\right)$  & $\left(2.43,2.37\right)$\tabularnewline
\hline 
$0.005$  & $\left(2.97,1.28\right)$  & $\left(3.12,1.52\right)$  & $\left(3.05,1.87\right)$\tabularnewline
\hline 
$0.008$  & $\left(3.33,1.13\right)$  & $\left(3.63,1.30\right)$  & $\left(3.55,1.59\right)$\tabularnewline
\hline 
\end{tabular}
\end{table}

\begin{table}
\caption{\label{tab:EK_eta_M_256}Values of$\left(E\left[K\right],\eta\right)$
for networks with $N=4,5,6$ sensors and source sequence length of$M=256$
bits.}

\centering%
\begin{tabular}{|c||c|c|c|}
\hline 
\multicolumn{1}{|c||||||}{$f_{d}T_{s}$} & $N=4$  & $N=5$  & $N=6$\tabularnewline
\hline 
$0.002$  & $\left(3.06,1.27\right)$  & $\left(3.25,1.50\right)$  & $\left(3.31,1.77\right)$\tabularnewline
\hline 
$0.005$  & $\left(3.62,1.06\right)$  & $\left(4.05,1.19\right)$  & $\left(4.23,1.37\right)$\tabularnewline
\hline 
$0.008$  & $\left(3.83,0.99\right)$  & $\left(4.46,1.07\right)$  & $\left(4.78,1.20\right)$\tabularnewline
\hline 
\end{tabular}
\end{table}

\section{\label{sec:Conclusion}Conclusion}

We analyzed a WSN where source-sensor channels are modeled as quantized
correlated fading channels (Gilbert-Elliott channels). We proposed
a two-phase transmission scheme where at the first phase compressed
channel state information sequences are transmitted to the FC and
a subset of sensors are selected to transmit their observations to
the FC at the second phase. Also, we analytically derived the probability
distribution of the size of the selected subset and the expected value
of this subset size. We presented simulation results to assess the
accuracy of our analyses. We defined an efficiency factor for our
proposed algorithm and evaluated this factor for several channel conditions
and network setups. In most cases our proposed two-phase algorithm
showed a superior power efficiency compared to a conventional one-phase
transmission scheme over slow-fading channels.

\section*{}


\begin{thebibliography}{10}
\bibitem{Survey_WSN}I. F. Akyildiz, W. Su, Y. Sankarasubramaniam,
and E. Cayirci, ``A survey on sensor networks,'' \textit{IEEE Communications
Magazine} (2002), Pages: 102-114.

\bibitem{singlenode1-1}Chin-Liang Wang, Syue-Ju Syue, ``An Efficient
Relay Selection Protocol for Cooperative Wireless Sensor Networks,''
\textit{Wireless Communications and Networking Conference}, 2009.
WCNC 2009. pp.1-5, April 2009.

\bibitem{singlenode2-1}Zorzi, M.; Rao, R.R., ``Geographic random
forwarding (GeRaF) for ad hoc and sensor networks: multihop performance,''
\textit{Mobile Computing, IEEE Transactions on }, vol.2, no.4, pp.337,348,
Oct.-Dec. 2003

\bibitem{Jointpoweralloc}X.J. Zhang and Y. Gong, ``Joint power allocation
and relay positioning in multi-relay cooperative systems'', \textit{IET
Commun.} (2009), vol. 3, Issue: 10, Pages:1683-1692.

\bibitem{sola_ WSN}Ranhotigamage, C., Mukhopadhyay, S.C., ``Field
Trials and Performance Monitoring of Distributed Solar Panels Using
a Low-Cost Wireless Sensors Network for Domestic Applications,''
\textit{Sensors Journal, IEEE} , vol.11, no.10, pp.2583,2590, Oct.
2011

\bibitem{target_WSN}Mahmut Karakaya, Huirong Qi, ``Target detection
and counting using a progressive certainty map in distributed visual
sensor networks'' \textit{International Conference on Distributed
Smart Cameras,} ICDSC 2009, Pages: 1-8.

\bibitem{petro_WSN}Zhang Ke, Li Yang, Xliao Wang-hui, Suh Heejong,
``The Application of a Wireless Sensor Network Design Based on ZigBee
in Petrochemical Industry Field'' \textit{International Conference
on Intelligent Networks and Intelligent Systems}, ICINIS 2008, Pages:
284-287.

\bibitem{Energy_Survey_WSN}G. Anastasi, M. Conti, M. Di Francesco,
A. Passarella, ``Energy conservation in Wireless Sensor Networks:
A Survey,'' \textit{Elsevier Ad Hoc Networks,} 7 (2009) 537-568.

\bibitem{Energy_aware_microsensor}V. Raghunathan, C. Schurghers,
S. Park, M. Srivastava, ``Energy-aware wireless microsensor networks,''
\textit{IEEE Signal Processing Magazine} (2002) Pages: 40\textendash{}50.

\bibitem{WSN_integrated}G. Pottie, W. Kaiser, ``Wireless integrated
network sensors,'' Communication of ACM 43 (5) (2000) 51\textendash{}58.

\bibitem{Gilbert}E.N. Gilbert, ``Capacity of a burst-noise channel,\textquotedblright{}
\textit{Bell Syst. Tech. J.}, vol. 39, no. 9, pp. 1253\textendash{}1265,
Sept. 1960.

\bibitem{Elliott}E.O. Elliott, ``Estimates of error rates for codes
on burst-noise channels,\textquotedblright{} \textit{Bell Syst. Tech.
J.}, vol. 42, no. 5, pp. 1977\textendash{}1997, Sept. 1963.

\bibitem{Sadeghi_Fading_Review}P. Sadeghi, R. A. Kennedy, P. B. Rapajic,
and R. Shams, ``Finite-State Markov Modeling of Fading Channels,''
\textit{IEEE Signal Processing Magazine,} Sep. 2008, pp. 57-80.

\bibitem{Willing08} A. Willing, ``Recent and emerging topics in
wireless industrial communications: A selection,'' \textit{IEEE Transactions
on Industrial Informatics,} vol. 4, n0. 2, pp. 102-124, May 2008.

\bibitem{Howard06}Sheryl L. Howard, Christian Schlegel, and Kris
Iniewski, ``Error control coding in low-power wireless sensor networks:
When is ECC energy efficient,'' \textit{EURASIP Journal on Wireless
Communications and Networking}, pages 1\textendash{}14, 2006.

\bibitem{Islam10}M. R. Islam ``Errorcorrection codes in wireless
sensor network: An energy aware approach,'' \textit{Int. J. Comput.Inf.
Eng.}, vol. 4, no. 1, pp.59 -64 2010

\bibitem{Abedi11}Abedi, A., ``Power-efficient-coded architecture
for distributed wireless sensing,'' \textit{Wireless Sensor Systems,
IET} , vol.1, no.3, pp.129,136, September 2011

\bibitem{FEC-adaptive-WSN}{]} K. Yu, F. Baracy, M. Gidlundz and J.
Akerbergz, ``Adaptive forward error correction for best effort wireless
sensor networks,\textquotedblright{} \textit{IEEE International Conf.
on Communications (ICC)}, Ottawa ,June 2012, pp. 7104-7109.

\bibitem{Jakes74}W. C. Jakes, \textit{Microwave Mobile Communications},
New York: IEEE Press, 1974.

\bibitem{Bonferroni36}C. Bonferroni, ``Teoria statistica delle classi
e calcolo delle probabilità,'' \textit{Pubbl. d. R. Ist. Super. di
Sci. Econom. e Commerciali di Firenze} (in Italian), vol. 8, Pages:
1\textendash{}62, 1936.\end{thebibliography}
\end{document}